\documentclass[11pt]{amsart}
\usepackage{geometry}                
\usepackage{graphicx}
\usepackage{amssymb}
\usepackage{epstopdf}
\title{Ocean Mass Redistribution \& Corresponding Frequency Offsets for Precise Timing Applications}
\date{April 4, 2012}
\author{Scott Czopek\\ 
	     \\ April 4, 2012 \\
	     \\  
	     NIST, Time \& Frequency Division \\
              \\ 
              scott.czopek@gmail.com}

\RequirePackage{fix-cm}
\usepackage{graphicx}
\usepackage{url}
\usepackage{amsmath}
\usepackage{placeins}

\begin{document}
\maketitle

\begin{abstract}
The relativistic timing affects of tidally redistributed ocean mass are investigated.  The Sun, Moon, and Earth hurl through space, and their gravitational fields cause the tides which is a slight redistribution of ocean mass.  This redistributed mass perturbs the Earth's gravitational potential affecting atomic clocks.  The magnitude of this perturbation will be quantified, correcting for this effect.

The predicted fractional frequency offset $(10^{-19})$ is too small to be detected, but this effect may become visible in the future.

\end{abstract}

\section{Introduction}
\label{intro}
The purpose of this paper is to investigate whether or not current time transfer techniques can detect a relativistic frequency shift, which would result from the tidal redistribution of ocean mass.  To answer this question software was written to predict how large this offset is.

The tides redistribute ocean mass across the surface of the Earth, from the South China Sea to the Bering Strait.  Millions of metric tons of water are being sloshed around, and this perturbs Earth's gravitational field.  This change has been detected\cite{GRACEtellus} \footnote{The twin GRACE satellites detect small changes in the Earth's gravitational field.  The oceans mass distribution is inferred from that.  What is done here is to infer a small change in Earth's field from a predicted ocean mass distribution.}, but is this change large enough to affect the precise transfer of time?

Gravitation can effect time transfer.  The gravitational potential difference alters the photons' frequency, and changes the rate at which a clock beats.  this perturbation can be calculated by computing tidal heights and representing them with spherical harmonics.

\section{Fractional Frequency Error Estimate}
\label{sec:1}
How large could this effect be?  Think of where the  frequency offset would be the largest.  This effect only appears when two distant locations compare their timing activities.  The effect is not observed locally.  There are two locations for where the potential difference could be the largest.  The largest potential difference could be between two locations on the ground where the potential is the most positive and the most negative \footnote{The potential can assume both negative and positive signs, because of low tides.  When the ocean recedes beneath sea level the missing water is treated as negative mass.}, or between the Earth's surface and a satellite.

The frequency difference will be greatest between two points on the ground.  If the maximum potential can be estimated, then assume that somewhere on the Earth's surface the potential is equally negative.  This mirror anti-image will introduce a factor of 2 into the estimation.  Remember, the difference between a positive number and a negative number is greater than the difference between that same positive number and zero.  The potential difference will be greatest between two points on the ground, rather than a point on the ground and a satellite far off in space.

This estimation will provide the maximum potential difference, which immediately provides the maximum fractional frequency offset.  While values may be smaller on some parts of the Earth, or up in space off the ground this larger number is more interesting.  It provides an idea of how large this effect will be and answer whether or not this effect can be measured with available technology.

To estimate how large this perturbation is going to be, a sum of spherical harmonics is considered.  Each spherical harmonic also has an associated radial component.  This sum is written as $\sum Y_l^m \frac{1}{r^{l+1}}$.  First the radial components will be approximated, and then the angular components.  Finally the proper coefficients will be attached to the summation.

Each solution to Poisson's Equation is a spherical harmonic associated with a radial component.  These associated radial components can be eliminated if the measurements are performed at sea level.  This is fine, because to detect this tiny fractional frequency offset it will have to be measured with some fancy equipment.  A large time and frequency lab would make this measurement, and these large time and frequency labs are located at sea level \footnote{The exception is NIST (National Institute of Standards and Technology), which houses a prominent timing lab, located in the Rocky Mountains (Boulder CO).  However, since the estimate strives to provide the largest possible fractional frequency offset this one exception remains just an exception and not a counter-example.}.

Sea level by definition is located at the Earth's surface.  And when these spherical harmonics and radial components are evaluated at the EarthÕs surface, all the radial parts collapse to $\frac{a^l}{r^{l+1}} = \frac{1}{a}$, where $a$ equals the Earth's radius.  This handles the radial components of $\sum Y_l^m \frac{1}{r^{l+1}}$.

Let's now turn our attention to the angular components of this equation $(\sum Y_l^m \frac{1}{r^{l+1}})$.  Each $Y_l^m$ is bounded by $\approx 1$, so set these to 1.  This handles the angular components of $\sum Y_l^m \frac{1}{r^{l+1}}$.

With $\sum Y_l^m \frac{1}{r^{l+1}}$'s radial components set to $\frac{1}{a}$ and the angular components set to 1, all that is left is to choose coefficients for each term in the series.  The terms need to represent how much water is being sloshed around by the tides.  So although each term's coefficient does not equal the total amount of water that is being sloshed around, all the coefficients together must equal this total.  Multiplying this sum by G turns it into a description of the gravitational potential.  By including one more coefficient $(\frac{1}{c^2})$ to the gravitational potential, the goal is achieved - the fractional frequency offset is calculated.

But how much water is being sloshed around by the tides?  The total displaced mass is a thin shell of water that covers the surface of the Earth.  The thickness of this shell is the average tidal height far out at sea - 10 cm.  The shell's density is the density of sea water - 1021 $\frac{\text{kg}}{\text{m}^3}$.  The height and the density provide a surface density of 102 $\frac{\text{kg}}{\text{m}^2}$.  Multiplying by the surface area of the Earth provides the mass  - $4\pi R^2 \sigma = 5.2 \times 10^{16}$ kg.  

Now to combine $4\pi \sigma R^2$ with the approximate radial component $(\frac{1}{R})$ and the approximate angular component $1$.  Finally adding G to transform it into a description of gravity.  This expression works out to be $4\pi \sigma R G$ Ð or $0.5 \frac{\text{m}^2}{\text{s}^2}$.

To compute the fractional frequency offset divide by $c^2$ \cite{lrrNeil}.  This gives a fractional error of $10^{-18}$.  This fractional error is measureable with cutting edge time keeping devices, but is much smaller than the error present in current time transfer techniques.  To detect this offset two clocks would have to be placed very far apart.  One clock would be located where there is a high tide and a high potential, and the other clock would be located where there is a low tide and a low potential.  The potential difference would result in a small measurement error in the time link between the two clocks.  However the fractional uncertainty $(10^{-18})$ is a thousand times smaller than the fractional uncertainty of the time link $(10^{-15})$ \cite{TFcompar}.  Several fractional frequency offsets will be calculated after a description of the software is given.

\section{Methods}
\label{sec:2}
The objective was to create a software package that calculates the ocean's tidal mass distribution, use that distribution as input, and output the fractional frequency offset at any latitude, longitude, or altitude. Relativistic perturbation theory is used to perform some of the calculations,  Avivo's FES 2004 (Finite Element Solution) tide prediction software is used, as is Mangle\cite{Mangle} (a program by Molly Swanson). 

``Relativity in the Global Positioning System"\cite{lrrNeil} fully describes how to translate a weak field potential into a fractional frequency offset that can be applied to an atomic clock.  This paper makes clear that fractional frequency offsets add linearly, so this additional offset can simply be added to the existing offsets.

To calculate the fractional frequency offset the software tools FES 2004 and Mangle are used.  FES 2004 is a tidal prediction program made available through Aviso.  It will provide the tidal height at a given location and time, which provides a numerical model for ocean mass redistribution.  This numerical model uses the tide height seen by a tide gauge for all calculations.  (The solid Earth loading is subtracted out.) 

Specifically this numerical model uses a mesh grid.  This mesh grid covers the Earth's surface and represents the tidal ocean mass redistribution.  From this mass distribution a perturbation is calculated.  Spherical harmonics were then fitted to the data using Mangle\cite{Mangle}, a software program written by Molly Swanson.  Mangle takes a surface density mask for a sphere and returns the spherical harmonic representation for that surface \footnote{Mangle \cite{Mangle} was originally written to process astronomical survey pixel maps and translate the surveys into spherical harmonics.  }.

Using these tools in tandem creates a software package that allows the perturbation to the Earth's potential to be evaluated at various locations, and the fractional frequency offset to be calculated.

\section{Results}
\label{sec:3}

$\text{  }$
\begin{figure}
	\begin{center}
		\includegraphics[trim=100 100 0 80, clip, scale=0.353]{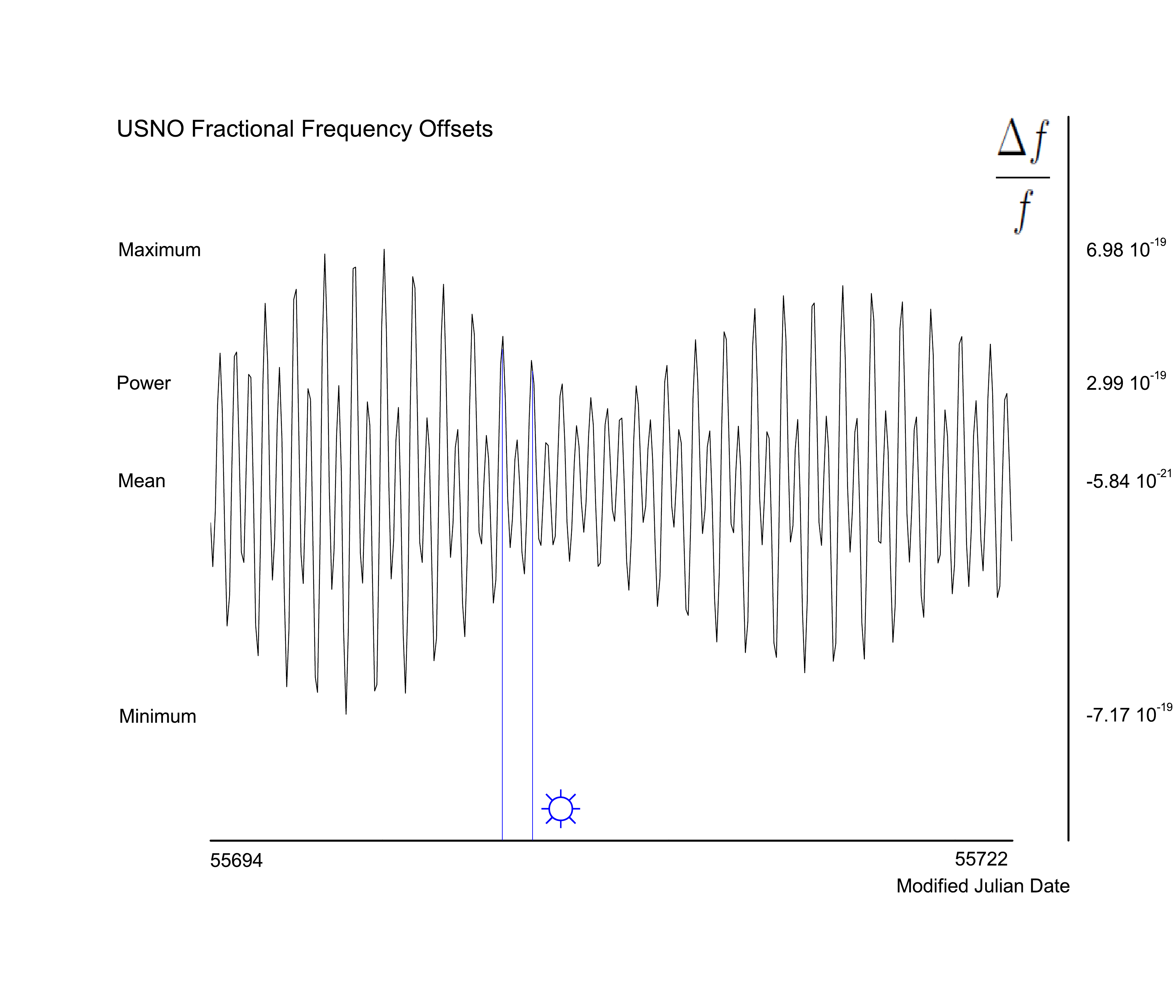}
		\includegraphics[trim=60 80 0 80, clip, scale=0.353]{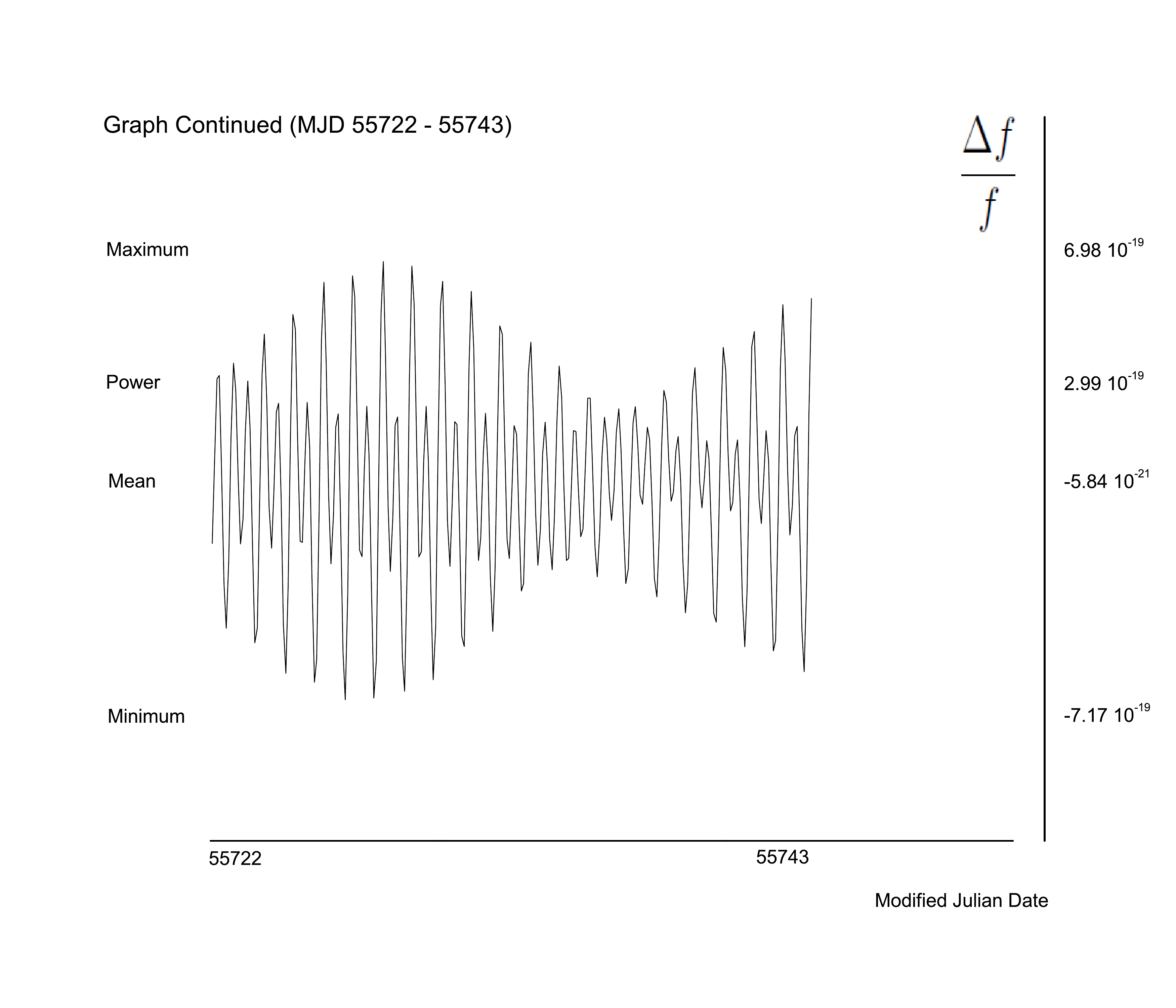}
	\end{center}
	\caption{USNO's predicted atomic clock's fractional frequency offset.  Redistributed tidal mass causes this offset.  05/13/2011 - 06/30/2011.  The two solid lines denote the width of a single day.  These fractional frequency offsets are typically $\sim 10^{-19}$.}
	\label{fig:USNOFracFreqOffsets}
\end{figure}

\begin{figure}
	\begin{center}
		\includegraphics[trim=91 0 103 0, clip, scale=0.5]{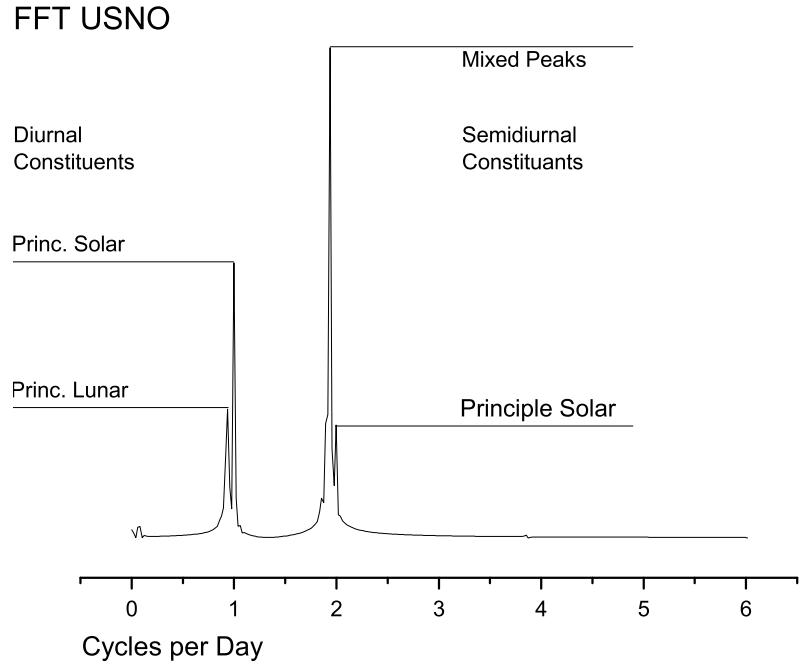}
	\end{center}
	\caption{This fourier transform lends confidence to the fractional frequency offset predictions.  Common tidal peaks are observed, which are related to the Moon and the Sun.  As expected, these peaks are located around one day and twelve hours.  Since the fourier transform plots for the other predictions show nearly the same thing, they are omitted.}
	\label{fig:USNOFFT}
\end{figure}

\begin{figure}
	\begin{center}
		\includegraphics[trim=100 80 10 100, clip, scale=0.375]{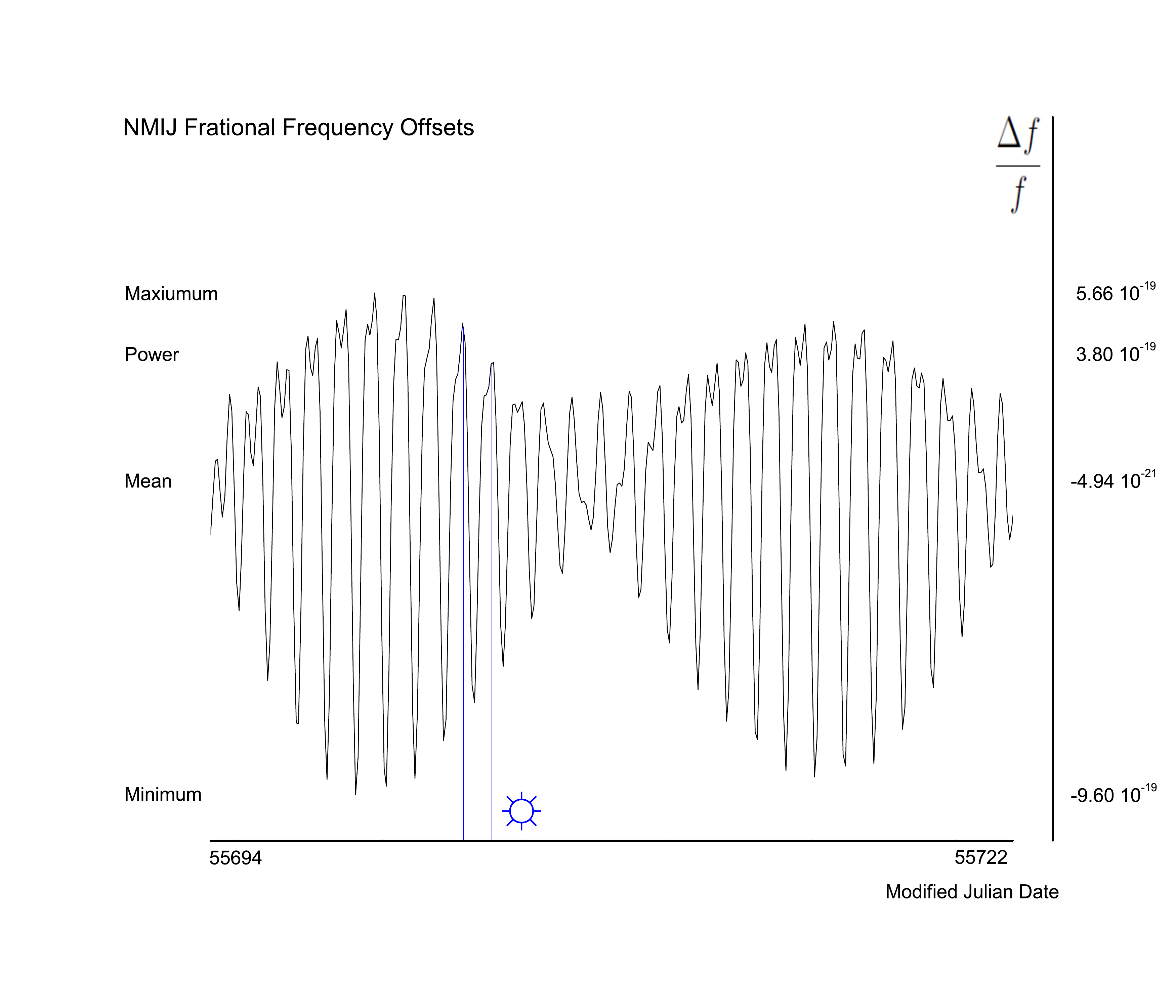}
		\includegraphics[trim=85 80 10 100, clip, scale=0.375]{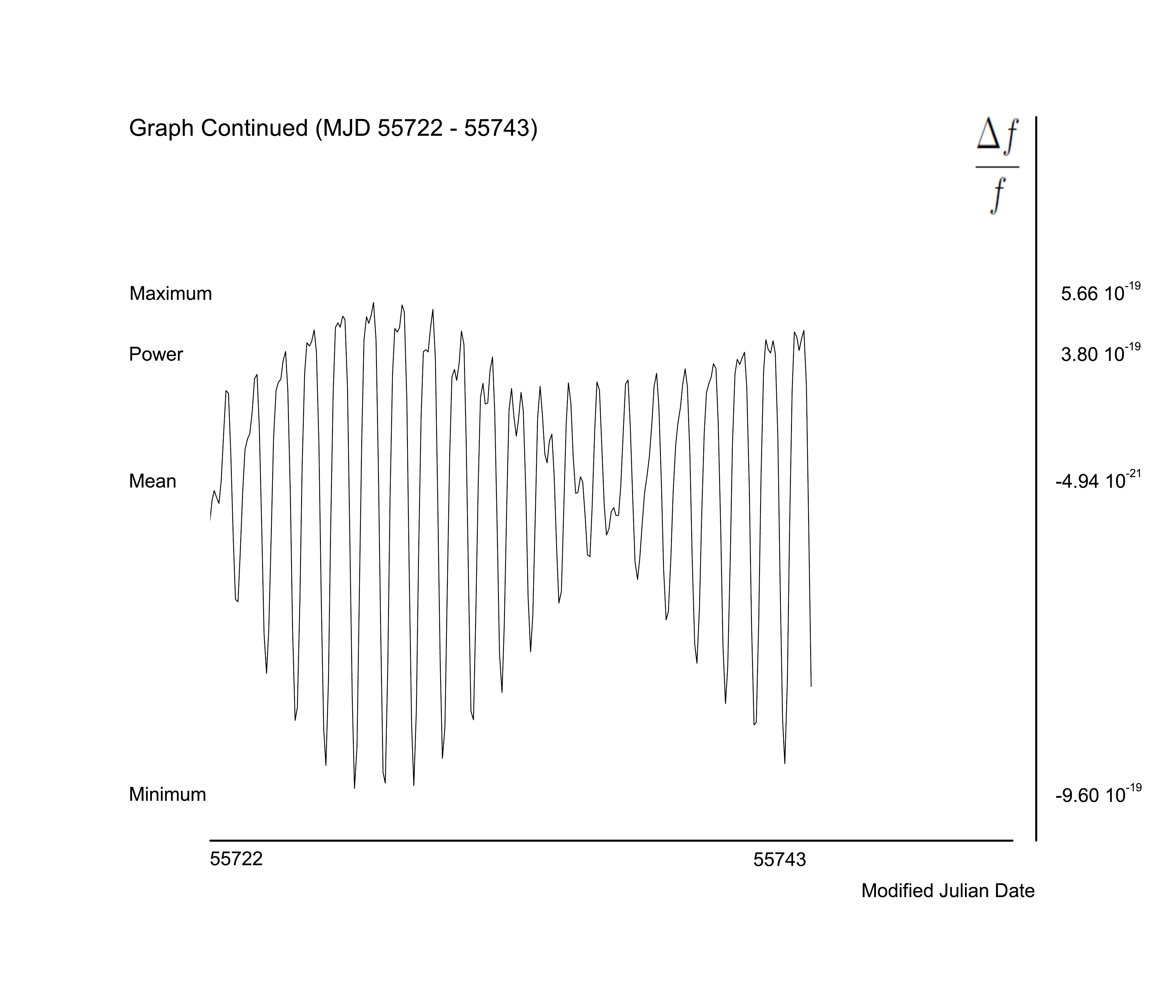}
	\end{center}
	\caption{\noindent This same type of prediction that was performed for the USNO, except that the fractional frequency offset for the NMIJ (National Meterological Institute of Japan) is predicted instead.  The MJD (Modified Julian Date) ranges from 05/13/2011 to 06/31/2011.  The two blue lines represent the width of a single day.  The magnitude of these frequency offsets are almost the same every where around the world $(\sim 10^{-19})$ - from America to Germany, to Japan.}
	\label{fig:NMIJFracFreqOffsets}
\end{figure}


\begin{figure}
	\begin{center}

		\vspace{20 pt}
		\large{\sffamily{Fractional Frequency Offset Statistics for National Labs}}\hspace{3 pt}
		\vspace{3 pt}
		\includegraphics[trim=90 620 206 100, clip, scale=1.0]{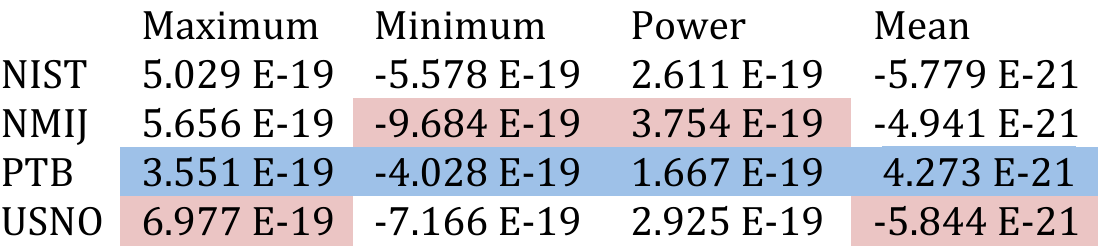}
		\fontfamily{sansserif}
		\selectfont
	\end{center}
	\caption{PTB (Physikalisch-Technische Bundesanstalt) experiences a surprisingly low level perturbation.  It has the least extreme statistics of the four labs examined.  These low level perturbations are highlighted in blue.  In contrast the most extreme statistics are highlighted in red.  Not surprisingly the most extreme numbers occur at national timing laboratories located by the sea shore (NMIJ - Japan, USNO - Washington D.C.).  This is where the tides will cause the largest effect.}
	\label{fig:TableExtremeValues}
\end{figure}

\FloatBarrier
\section{Discussion}
\label{sec:4}
The findings indicate that the fractional frequency offset is roughly the same magnitude that is estimated.  The estimation suggested that the fractional error would be $10^{-18}$ and the numerically predicted figure is $10^{-19}$.  Although this offset is small does it accumulate over time?  No.

The error might accumulate during the first half of the cycle, where all the fractional frequency errors share the same sign$( + \text{ or } -)$.  Since the beat frequencies of this phenomena are tied to the Earth's orbit around the Moon and the Sun this means that the half cycles' lengths are nearly 6 hours or nearly 12 hours.  The max peak height is a fractional frequency offset of $9 \times 10^{-19}$, so after integrating for the greatest length of time (12 hours) a few fempto-second error accumulates.  This error might be detectable if a precise time signal were to be translated from one optical ion clock in one part of the globe to another, but this comparison is not possible because of current limitations for time and frequency transfer.  So this error is not important right now, but might become important in the near future.

Since the mean value is not zero, there could be an underlying non-zero offset.  What is more likely is that this non-zero mean results from adding up a partial cycle at the end of the simulation.  The average value is 100 times smaller than the signal's root mean square power and the data set contains $\sim$ 600 points.  This means that the data set contains 49 full cycles, plus a little bit of the $50^{\text{th}}$ cycle too.  The full cycles balance out to zero, and the incomplete cycle is divided by the number of points in the data set.  The series' sum is reduced from $10^{-19}$ to $10^{-21}$ the data set's average value.  This small mean value that is very close to zero is not something to worry about, because the partial cycle error masks any underlying non-zero mean value trend.

\section{Conclusion}
\label{sec:5}
A fractional frequency offset of $10^{-19}$ is small, and currently it is not measureable.  It is $10^2$ times smaller than the best clock, and $10^4$ times smaller than the world's best time transfer link.  This error does not need to be worried about in the near future.  But perhaps in the future small geo-physical effects will need to be accounted for to keep delivering the maximum available precision provided by the world's best time standards.

\section{acknowledgements}
\label{sec:6}
Here is where I tip my hat to some very special people.  To Dr. Andrew Hamilton for his unlimited guidance and great concern for my career.  To Dr. Neil Ashby for his technical assistance.  To Dr. Tom O'Brian for sponsoring me as a guest researcher at NIST.  Thank you all so very much.



\end{document}